\documentclass[]{aa}
\usepackage{graphicx}
\usepackage{natbib}
\usepackage{multirow}
\usepackage{tabularx}
\usepackage{amssymb}
\usepackage{amsmath}
\usepackage{euscript}
\usepackage{bm}
\usepackage{pdflscape}
\usepackage{tikz}
\usepackage{pifont}
\usepackage[super]{nth}

\bibpunct{(}{)}{;}{a}{}{,} 

\newcommand*{\tran}{^{\mkern-1.5mu\mathsf{T}}}

\begin{document}

   \title{Estimating activity cycles with probabilistic methods I. \\Bayesian Generalised Lomb-Scargle Periodogram with Trend}
   \authorrunning{N. Olspert et al.}
   \titlerunning{Estimating activity cycles with probabilistic methods}

   \author{N. Olspert \inst{1} \and 
    J. Pelt \inst{2} \and
   	M. J. K\"apyl\"a\ \inst{3,1} \and 
   	J. Lehtinen \inst{3,1}}
   \offprints{N. Olspert\\
          \email{nigul.olspert@aalto.fi}
	}
\institute{
    ReSoLVE Centre of Excellence, Aalto University, Department of Computer Science, PO Box 15400, FI-00076 Aalto, Finland
\and Tartu Observatory, 61602 T\~{o}ravere, Estonia
\and Max-Planck-Institut f\"ur Sonnensystemforschung,
Justus-von-Liebig-Weg 3, D-37077 G\"ottingen, Germany
}

\date{Received / Accepted}

\abstract{Period estimation is one of the central topics in astronomical time series
  analysis, where data is often unevenly sampled. Especially challenging are studies of
  stellar magnetic cycles, as there the periods looked for are of the order of the same length
  than the datasets themselves. The datasets often contain trends, the origin of which is either
  a real long-term cycle or an instrumental effect, but these effects cannot be reliably separated,
  while they can lead to erroneous period determinations if not properly handled.}
  {In this study we aim at developing a method that can handle the trends properly, and by performing
  extensive set of testing, we show that this is the optimal procedure when contrasted with methods that do 
  not include the trend directly to the model.
  The effect of the form of the noise (whether constant or heteroscedastic) on the results is also investigated.}
{We introduce a Bayesian Generalised Lomb-Scargle Periodogram with Trend (BGLST), which is a
	probabilistic linear regression model using Gaussian priors for the coefficients of the fit
	and uniform prior for the frequency parameter.
}
{We show,
  using synthetic data,
  that when there is no prior information on whether and to what extent the true model
  of the data contains a linear trend, the introduced BGLST method is preferable to
  the methods which either detrend the data or 
  leave the data undetrended before fitting the periodic model.
  Whether to use noise with different than constant variance in the model depends on the density of the data sampling as well as on the true noise type of the process.
  }
{}

 
\maketitle

\section{Introduction}
In the domain of astronomical data analysis the task of period estimation from unevenly spaced time series has been
relevant topic for many decades. Depending on the context, the term ``period'' can refer to e.g. rotational 
period of the star, orbiting period of the star or exoplanet, or the period of the activity cycle of the star. 
Knowing the precise value of the period is often very important as many other physical quantities are
dependent on it. For instance, in the case of active late-type stars with outer convection
zones, the ratio between 
rotational period and cycle period can be interpreted as
a measure of the dynamo efficiency.
Therefore, measuring this ratio gives us crucial information of the generation
of magnetic fields in stars with varying activity levels.
In practice, both periods are usually estimated through photometry and/or spectrometry of the star.

For the purpose of analysing unevenly spaced time series, many different methods
have been developed over the years. 
Historically one of the most used ones is the Lomb-Scargle (LS) periodogram \citep{Lomb1976, Scargle1982}. 
Statistical properties of the LS and other alternative periodograms have been
extensively studied and it has become a
well-known fact that the interpretation of the results of any spectral analysis
method takes a lot of effort.
The sampling patterns in the data together with the finiteness of the time span
of observations lead to a
multitude of difficulties in the period estimation. 
One of the most pronounced difficulties is the emergence of aliased peaks (spectral leakage)
\citep{Tanner1948, Deeming1975, Lomb1976, Scargle1982, Horne1986}.
In other words one can only see a distorted spectrum as the 
observations are made at discrete (uneven) time moments and during a finite time.
However, algorithms for eliminating spurious peaks from spectrum have been developed \citep{Roberts1987}.
Clearly, different period estimation methods tend to perform differently depending on the dataset.
For comparison of some of the popular methods see \citet{Carbonell1992}.

One of the other issues in spectral estimation arises when the true mean of the measured signal is not known.
The LS method assumes a zero-mean harmonic model with constant noise variance,
so that the data needs to be centered in the observed values before doing the analysis. While every dataset can be
turned into a dataset with a zero mean, in some cases, due to pathological sampling
(if the empirical mean\footnote{We use the terms ``empirical mean'' and ``empirical trend'' 
when referring to the values obtained by correspondingly fitting a constant or a 
line to the data using linear regression.}
and the true mean differ significantly) this procedure can lead
to incorrect period estimates. In the literature
this problem has been addressed extensively and several generalizations of the periodogram, which are
invariant to the shifts in the observed value, have been proposed.
Such include the method of
\citet{Ferraz-Mello1981}, who used Gram-Schmidt orthogonalization of the constant, cosine
and sine functions in the sample domain. The power spectrum is then defined as the square
norm of the data projections
to these functions. In \citet{Zechmeister2009} the harmonic model of the LS periodogram is directly
extended with the addition of a constant term, which has become known as the Generalised Lomb-Scargle
(GLS) periodogram. Moreover, both of these studies give the formulations allowing nonconstant noise variance. 
Later, using a Bayesian approach for a model with harmonic plus constant, it has been shown that
the posterior probability of the frequency, when using uniform priors, is very similar to the GLS
spectrum \citep{Mortier2015}. The benefit of the latter method is that the relative probabilities
of any two frequencies can be easily calculated. 
Usually in the models, likewise in the current study, the noise is assumed to be Gaussian and uncorrelated. 
Other than white noise models are discussed in \citet{Vio2010}
and \citet{Feng2017}.
For more thorough overview of important aspects in the spectral 
analysis please refer to \citet{VanderPlas2017}.

The focus of the current paper is on another yet unaddressed issue, namely the effect of a linear 
trend in the data to period estimation.
The motivation to tackle this question arose in the context of analysing the
Mount Wilson time series of chromospheric activity (hereafter MW)
in a quest to look for long-term activity cycles \citep{Olspert2017}, the length
of which is of the same order of
magnitude than the data set length itself.
In a previous study by \citep{Baliunas1995}
detrending was occassionally, but not systematically, used before the LS periodogram was calculated.
The cycle estimates from this study have been later used
extensively by many other studies, for example, to show the presence of different stellar
activity branches \citep[e.g.][]{SB99}.

We now raise the following question -- whether it is more optimal to remove
the linear trend before the period search 
or leave the data undetrended?
Likewise with centering, detrending the data before fitting the harmonic model may or may not lead to biased period estimates 
depending on the structure of the data. The problems arise either due to sampling effects or due to presence of very long periods 
in the data. Based on the empirical arguments given in Sect.~\ref{experiments}, we show that it is more preferable to 
include the trend component directly into the regression model instead of detrending the data 
{\it a priori} or leaving the data 
undetrended altogether. In the same section we also discuss the effects of noise model of the data to the period estimate.
We note that the question that we address in the current study is primarily relevant to the
cases where the number of cycles in the dataset is small.
Otherwise, the
long coherence time (if the underlying process is truly periodic) allows to nail down periods, even if there are uneven errors or a small trend.
High number of cycles allows to get exact periods even e.g. by cycle counting.

The generalised least squares spectrum allowing arbitrary components
(including linear trend) was first discussed 
in \citet{Vanicek1971} 
and more recently Bayesian approaches including trend 
component have been introduced in \citet{Ford2011}
and \cite{Feng2017}.

\section{Method}\label{harmonic_model}

Now we turn to the description of the present method, which is a generalization of the method proposed in 
\cite{Mortier2015} and a special case of the method developed in \cite{Feng2017}. 
The model in the latter paper allows noise to be correlated, which we do not consider in the current study.
One of the main differences between the models discussed here and the one in \cite{Feng2017} is that we do 
not use uniform, but Gaussian priors for the nuisance parameters (see below). 
This has important 
consequences in certain situations as discussed in Sect.~\ref{priors_discussion}.
We introduce a simple Bayesian linear regression model where besides the harmonic component we have
a linear trend with slope plus offset\footnote{The implementation of the method introduced in this paper can be found at \url{https://github.com/olspert/BGLST}}.
This is summarised in the following equation:
\begin{equation}\label{eq_harmonic_model}
y(t_i)=A\cos(2\pi f t_i - \phi) + B\sin(2\pi f t_i - \phi) + \alpha t_i + \beta + \epsilon(t_i),
\end{equation}
where $y(t_i)$ and $\epsilon(t_i)$ are the observation and noise at time 
$t_i$, $f=1/P$ is the frequency of the cycle, $A$, $B$, $\alpha$, $\beta$ and $\phi$ are
free parameters. Specifically, $\alpha$ is the slope and $\beta$ the offset (y-intercept).
Usually $A$, $B$, $\alpha$, $\beta$ are called the nuisance parameters.
As noted, we assume 
that the noise is Gaussian and independent between any two time moments, but we 
allow its variance to be time dependent. For parameter inference we use a Bayesian model, 
where the posterior probability is given by 
\begin{equation}\label{eq_posterior}
\begin{aligned}
p(f,\bm{\theta}|D) &\propto p(D|f,\bm{\theta})p(f, \bm{\theta}),
\end{aligned}
\end{equation}
where $p(D|f,\bm{\theta})$ is the likelihood of the data, $p(f, \bm{\theta})$ is the prior probability of 
the parameters,
where for convenience, we have grouped the nuisance parameters under the vector $\bm{\theta} = \left[A, B, \alpha, \beta\right]\tran$. 
Parameter 
$\phi$ is not optimized, but set to a frequency dependent value simplifying 
the inference such that cross terms with cosine and sine components 
will vanish (see Sect.~\ref{harmonic_model_details}). The likelihood is given by
\begin{equation}\label{eq_likelihood}
p(D|f,\bm{\theta})=\left(\prod_{i=1}^{N}\frac{1}{\sqrt{2\pi}\sigma_i}\right)\exp\left(-\frac{1}{2}\sum_{i=1}^{N}\frac{\epsilon_i^2}{\sigma_i^2}\right),
\end{equation}
where $\epsilon_i=\epsilon(t_i)$ and $\sigma^2_i$ is the noise variance at time moment $t_i$.
To make the derivation of the spectrum analytically tractable we take independent Gaussian priors for $A, B, \alpha, \beta$ and
a flat prior for the frequency $f$. This leads to
the prior probability given by
\begin{equation}\label{eq_prior}
p(f, \bm{\theta})=\mathcal{N}(\bm{\theta}|\bm{\mu}_{\theta}, \bm{\Sigma}_{\theta}),
\end{equation}
where $\bm{\mu}_{\theta}=\left[\mu_A, \mu_B, \mu_{\alpha}, \mu_{\beta}\right]\tran$ is the vector of prior means and $\bm{\Sigma}_{\theta}={\rm diag}(\sigma^2_A, \sigma^2_B, \sigma^2_{\alpha}, \sigma^2_{\beta})$ is the diagonal matrix of prior variances.

The larger the prior variances the less information is assumed to be known about the 
parameters. Based on what is intuitively meaningful, in all the calculations we have 
used the following values for the prior means and variances:
\begin{equation}\label{eq_priors}
\begin{aligned}
\mu_A&=0, \mu_B=0, \mu_{\alpha}=\alpha_{\rm slope}, \mu_{\beta}=\beta_{\rm intercept},\\
\sigma^2_A&=0.5\sigma^2_y, \sigma^2_B=0.5\sigma^2_y, \sigma^2_{\alpha}=\frac{\sigma^2_y}{\Delta T^2},\sigma^2_{\beta}=\sigma^2_y + \beta^2_{\rm intercept},
\end{aligned}
\end{equation}
where $\alpha_{\rm slope}$ and $\beta_{\rm intercept}$ are the slope and intercept estimated from linear 
regression, $\sigma^2_y$ is the sample variance of the data and $\Delta T$ is duration of the data. If one 
does not have any prior information about the parameters one could set the variances to infinity and drop the 
corresponding terms from the equations, but in practice to avoid
meaningless results with unreasonably large parameter values some regularization would be required
(see Sect.~\ref{priors_discussion}).

\subsection{Derivation of the spectrum}\label{harmonic_model_details}
Our derivation of the spectrum closely follows the key points presented in \cite{Mortier2015}.
Consequently, we 
use as identical notion of the variables as possible.
Using the likelihood and prior defined by Eqs.~(\ref{eq_likelihood}) and (\ref{eq_prior}), the posterior 
probability of the Bayesian model given by Eq.~(\ref{eq_posterior}) can be explicitly written as 
\begin{equation}\label{eq_posterior_explicit}
\begin{aligned}
p(f,\bm{\theta}|D) &\propto p(D|f,\bm{\theta})p(\bm{\theta})\\
&= p(D|f,\bm{\theta})\mathcal{N}(\bm{\theta}|\bm{\mu}_{\theta}, \bm{\Sigma}_{\theta})\\
&=\prod_{i=1}^{N}\frac{1}{\sqrt{2\pi}\sigma_i}\prod_{i=1}^{k}\frac{1}{\sqrt{2\pi}\sigma_{\theta_i}}e^{-\frac{1}{2}E},
\end{aligned}
\end{equation}
where
\begin{equation}
\begin{aligned}
E=\sum_{i=1}^{N}\frac{\epsilon_i^2}{\sigma_i^2}+\sum_{i=1}^{k}\frac{({\theta_i}-\mu_{\theta_i})^2}{\sigma^2_{\theta_i}}.
\end{aligned}
\end{equation}
Here $\theta_i,\ i=1,..,k,\ k=4$, denotes the $i$th element of $\bm{\theta}$, i.e. either $A$, $B$, 
$\alpha$, or $\beta$.
From Eqs.~(\ref{eq_harmonic_model}) and (\ref{eq_likelihood}) we see that the likelihood and therefore posterior 
probability for every fixed frequency $f$ is multivariate Gaussian w.r.t. the parameters $A$, $B$, $\alpha$ and 
$\beta$. In principle we are interested in finding the optimum for the full joint posterior probability density 
$p(f,\bm{\theta}|D)$, but as for every fixed frequency the latter distribution is a multivariate Gaussian we can 
first marginalise over all nuisance parameters, find the optimum for the frequency $f$ e.g. by doing grid search 
and later analytically find the posterior means and covariances of the other parameters. The marginal 
posterior distribution 
for the frequency parameter is expressed as
\begin{equation}\label{eq_posterior_marginal}
p(f|D) \propto \int p(D|f,\bm{\theta})\mathcal{N}(\bm{\theta}|\bm{\mu}_{\theta}, \bm{\Sigma}_{\theta}) d\bm{\theta}.
\end{equation}
Here the integrals are assumed to be taken over the whole range of parameter values, i.e. from $-\infty$ to $\infty$.
We introduce the following notations:
\begin{eqnarray}
w_i&=&\frac{1}{\sigma^2_i}, w_A=\frac{1}{\sigma^2_A}, w_B=\frac{1}{\sigma^2_B}, 
w_{\alpha}=\frac{1}{\sigma^2_{\alpha}}, w_{\beta}=\frac{1}{\sigma^2_{\beta}},\\
W&=&\sum_{i=1}^{N}w_i + w_{\beta}, Y = \sum_{i=1}^{N}w_iy_i + w_{\beta} \mu_{\beta},\\
C&=&\sum_{i=1}^{N}w_i\cos(2\pi f t_i-\phi),\\
S&=&\sum_{i=1}^{N}w_i\sin(2\pi f t_i-\phi),\\
T&=&\sum_{i=1}^{N}w_it_i,\\
\widehat{YC}&=&\sum_{i=1}^{N}w_iy_i\cos(2\pi f t_i-\phi) + w_{A} \mu_{A},\\
\widehat{YS}&=&\sum_{i=1}^{N}w_iy_i\sin(2\pi f t_i-\phi) + w_{B} \mu_{B},
\end{eqnarray}
\begin{eqnarray}
\widehat{CC}&=&\sum_{i=1}^{N}w_i\cos^2(2\pi f t_i-\phi) + w_{A},\\
\widehat{SS}&=&\sum_{i=1}^{N}w_i\sin^2(2\pi f t_i-\phi) + w_{B},\\
\widehat{TT}&=&\sum_{i=1}^{N}w_it_i^2+ w_{\alpha},\\
\widehat{YT}&=&\sum_{i=1}^{N}w_iy_it_i + w_{\alpha} \mu_{\alpha},\\
\widehat{TC}&=&\sum_{i=1}^{N}w_it_i\cos(2\pi f t_i-\phi),\\
\widehat{TS}&=&\sum_{i=1}^{N}w_it_i\sin(2\pi f t_i-\phi),\\
\widehat{YY}&=&\sum_{i=1}^{N}w_iy_i^2+w_A \mu_{A}^2+w_B \mu_{B}^2+w_{\alpha} \mu_{\alpha}^2+w_{\beta} \mu_{\beta}^2.
\end{eqnarray}
If the value of $\phi$ is defined such that the cosine and sine functions are
orthogonal \citep[for the proof see][]{Mortier2015}, namely
\begin{equation}
\phi=\frac{1}{2}\arctan\left[\frac{\sum_{i=1}^{N}w_i\sin(4\pi f t_i)}{\sum_{i=1}^{N}w_i\cos(4\pi f t_i)}\right],
\end{equation}
then we have
\begin{equation}\label{integrand}
\begin{aligned}
E &=
\widehat{CC}A^2-2\widehat{YC}A+2\alpha\widehat{TC}A+2\beta CA\\
&+\widehat{SS}B^2-2\widehat{YS}B+2\alpha\widehat{TS}B+2\beta SB\\
&+\widehat{TT}\alpha^2-2\widehat{YT}\alpha+2T\beta\alpha+W\beta^2-2Y\beta+\widehat{YY},
\end{aligned}
\end{equation}
where we have grouped the terms with $A$ on the first line, terms with $B$ on the second line and the rest of the 
terms on the third line.

In the following we will repeatedly use the formula of the following definite integral:
\begin{equation}\label{known_integral}
\int_{-\infty}^{\infty}e^{-ax^2-2bx}dx=\sqrt{\frac{\pi}{a}}e^{\frac{b^2}{a}}, \quad a > 0.
\end{equation}
To calculate the integral in Eq.~(\ref{eq_posterior_marginal}), we start by integrating first over $A$ and $B$. 
Assuming that 
$\widehat{CC}$ and $\widehat{SS}$ are greater than zero and applying Eq.~(\ref{known_integral}), 
we will get the solution for the integral containing terms with A:
\begin{equation}\label{int_A}
\begin{aligned}
\int_{-\infty}^{\infty}\exp\left(-\frac{\widehat{CC}A^2-2\widehat{YC}A+2\alpha\widehat{TC}A+2\beta CA}{2}\right)dA\\
=\sqrt{\frac{2\pi}{\widehat{CC}}}\exp\left(\frac{(\widehat{YC}-\alpha \widehat{TC}-\beta C)^2}{2\widehat{CC}}\right)\\
=\exp\left(\frac{-\alpha\widehat{YC}\widehat{TC}}{\widehat{CC}}
+\frac{\alpha^2\widehat{TC}^2}{2\widehat{CC}}
+\frac{\alpha\beta\widehat{TC}C}{\widehat{CC}}\right)\\
\cdot\exp\left(\frac{-\beta\widehat{YC}C}{\widehat{CC}}
+\frac{\beta^2 C^2}{2\widehat{CC}}\right)\\
\cdot \sqrt{\frac{2\pi}{\widehat{CC}}}\exp\left(\frac{\widehat{YC}^2}{2\widehat{CC}}\right).
\end{aligned}
\end{equation}
In the last expression we have grouped onto separate lines the terms with $\alpha$,
terms with $\beta$ not simultaneously containing $\alpha$, and constant terms.
Similarly for the integral containing terms with $B$:
\begin{equation}\label{int_B}
\begin{aligned}
\int_{-\infty}^{\infty}\exp\left(-\frac{\widehat{SS}B^2-2\widehat{YS}B+2\alpha\widehat{TS}B+2\beta SB}{2}\right)dB\\
=\sqrt{\frac{2\pi}{\widehat{SS}}}\exp\left(\frac{(\widehat{YS}-\alpha \widehat{TS}-\beta S)^2}{2\widehat{SS}}\right)\\
=\exp\left(\frac{-\alpha\widehat{YS}\widehat{TS}}{\widehat{SS}}
+\frac{\alpha^2\widehat{TS}^2}{2\widehat{SS}}
+\frac{\alpha\beta\widehat{TS}S}{\widehat{SS}}\right)\\
\cdot\exp\left(\frac{-\beta\widehat{YS}S}{\widehat{SS}}
+\frac{\beta^2 S^2}{2\widehat{SS}}\right)\\
\cdot \sqrt{\frac{2\pi}{\widehat{SS}}}\exp\left(\frac{\widehat{YS}^2}{2\widehat{SS}}\right).
\end{aligned}
\end{equation}
Now we gather the coefficients for all the terms with $\alpha^2$ 
from the last line of Eq.~(\ref{integrand}) (keeping in mind the factor -1/2) as well as from 
Eqs.~(\ref{int_A}),~(\ref{int_B}) into new variable $K$:
\begin{equation}
\begin{aligned}
K=\frac{1}{2}\left(-\widehat{TT}+\frac{\widehat{TC}^2}{\widehat{CC}}+\frac{\widehat{TS}^2}{\widehat{SS}}\right).
\end{aligned}
\end{equation}
We similarly introduce new variable L for the coefficients involving all terms with $\alpha$ from the same equations:
\begin{equation}
\begin{aligned}
L=\left(\widehat{YT}-\beta T + \frac{-\widehat{YC}\widehat{TC}+\beta\widehat{TC}C}{\widehat{CC}}+\frac{-\widehat{YS}\widehat{TS}+\beta\widehat{TS}S}{\widehat{SS}}\right).
\end{aligned}
\end{equation}
After these substitutions, integrating over $\alpha$ can again be accomplished with the help of 
Eq.~(\ref{known_integral}) and assuming $K<0$:
\begin{equation}\label{int_alpha}
\begin{aligned}
\int_{-\infty}^{\infty}\exp(K\alpha^2+L\alpha)=\sqrt{\frac{\pi}{-K}}\exp\left(\frac{L^2}{-4K}\right)\\
=\sqrt{\frac{\pi}{-K}}\exp\left(\frac{(M+N\beta)^2}{-4K}\right)\\
=\sqrt{\frac{\pi}{-K}}\exp\left(\frac{N^2\beta^2}{-4K}\right)\exp\left(\frac{2MN\beta}{-4K}\right)\exp\left(\frac{M^2}{-4K}\right),
\end{aligned}
\end{equation}
where
\begin{equation}
M=\widehat{YT}-\frac{\widehat{YC}\widehat{TC}}{\widehat{CC}}-\frac{\widehat{YS}\widehat{TS}}{\widehat{SS}}
\end{equation}
and
\begin{equation}
N=\frac{\widehat{TC}C}{\widehat{CC}}+\frac{\widehat{TS}S}{\widehat{SS}}-T.
\end{equation}
At this point what is left to do is the integration over $\beta$. To simplify things
once more, we gather the coefficients for all the terms with $\beta^2$ 
from Eqs.~(\ref{integrand}),~(\ref{int_A}),~(\ref{int_B}),~(\ref{int_alpha}) into new variable $P$:
\begin{equation}
P=\frac{C^2}{2\widehat{CC}}+\frac{S^2}{2\widehat{SS}}-\frac{W}{2}-\frac{N^2}{4K}.
\end{equation}
We similarly introduce new variable Q for the coefficients involving all terms with $\beta$ from the same equations:
\begin{equation}
Q=-\frac{\widehat{YC}C}{\widehat{CC}}-\frac{\widehat{YS}S}{SS}+Y-\frac{2MN}{4K}.
\end{equation}
With these substitutions we are ready to integrate over $\beta$ using again Eq.~(\ref{known_integral}) while assuming $P<0$:
\begin{equation}
\begin{aligned}
\int_{-\infty}^{\infty}\exp(P\beta^2+Q\beta)=\sqrt{\frac{\pi}{-P}}\exp\left(\frac{Q^2}{-4P}\right).
\end{aligned}
\end{equation}
After gathering all remaining constant terms from Eqs.~(\ref{integrand}),~(\ref{int_A}),~(\ref{int_B}),~(\ref{int_alpha}), we finally obtain:

\begin{equation}\label{prob_spec}
\begin{aligned}
p(f|D) \propto &\frac{2\pi^2}{\sqrt{(\widehat{CC}\widehat{SS}KP)}}\\
&\cdot 
\exp\left(\frac{\widehat{YC}^2}{2\widehat{CC}}+\frac{\widehat{YS}^2}{2\widehat{SS}}-\frac{M^2}{4K}-\frac{Q^2}{4P}-\frac{\widehat{YY}}{2}\right).
\end{aligned}
\end{equation}

For the purpose of fitting the regression curve into the data one would be interested in obtaining the 
expected
values for the nuisance parameters $A, B, \alpha, \beta$. This can be easily done after fixing the frequency 
to its optimal value using Eq.~(\ref{prob_spec}) with grid search, and noticing that 
$p(\bm{\theta}|D,f_{\rm opt})$ is a multivariate Gaussian distribution. In the following we list the 
corresponding posterior means of the parameters\footnote{This is similar to Empirical Bayes approach as we use the point estimate for $f$.}: 

\begin{eqnarray}
\mu_{\beta} &=&-\frac{Q}{2P},\\
\mu_{\alpha}&=&-\frac{N\mu_{\beta}+M}{2K},\\
\mu_A&=&\frac{\widehat{YC}-\widehat{TC}\mu_{\alpha}-C\mu_{\beta}}{\widehat{CC}},\\
\mu_B&=&\frac{\widehat{YS}-\widehat{TS}\mu_{\alpha}-S\mu_{\beta}}{\widehat{SS}}.
\end{eqnarray}
Formulas for the full covariance matrix of the parameters as well as for the posterior
predictive distribution, assuming a model with
constant noise variance, can be found in \citet[Chapter~7.6]{Murphy}.
The posterior predictive distribution, however, in our case does not include the uncertainty 
contribution from the frequency parameter.

If we now consider the case when $\widehat{SS}=0$ \citep[for more details see][]{Mortier2015}, we see that also 
$S=0$, $\widehat{YS}=0$ and $\widehat{TS}=0$. Consequently the integral for B is proportional to a constant. We 
can define the analogues of the constants K through Q for 
this special case as
\begin{eqnarray}
K_C&=&\frac{1}{2}\left(-\widehat{TT}+\frac{\widehat{TC}^2}{\widehat{CC}}\right),\\
L_C&=&\left(\widehat{YT}-\beta T + \frac{-\widehat{YC}\widehat{TC}+\beta\widehat{TC}C}{\widehat{CC}}\right),\\
M_C&=&\widehat{YT}-\frac{\widehat{YC}\widehat{TC}}{\widehat{CC}},\\
N_C&=&\frac{\widehat{TC}C}{\widehat{CC}}-T,\\
P_C&=&\frac{C^2}{2\widehat{CC}}-\frac{W}{2}-\frac{N_C^2}{4K_C},\\
Q_C&=&-\frac{\widehat{YC}C}{\widehat{CC}}+Y-\frac{2M_CN_C}{4K_C}.
\end{eqnarray}
Finally, we arrive at the expression for the data likelihood 
\begin{equation}\label{prob_spec_special}
\begin{aligned}
p(f|D) \propto &\frac{2\pi^2}{\sqrt{(\widehat{CC}K_CP_C)}}\\
&\cdot
\exp\left(\frac{\widehat{YC}^2}{2\widehat{CC}}-\frac{M_C^2}{4K_C}-\frac{Q_C^2}{4P_C}-\frac{\widehat{YY}}{2}\right).
\end{aligned}
\end{equation}
Similarly we can handle the situation when $\widehat{CC}=0$. In the derivation we also assumed that $K<0$ and $P<0$. 
Our experiments with test data showed that the condition for $K<0$ was always satisfied, but occasionally P 
obtained values zero and probably due to numerical rounding errors also very low positive values.
These special cases we handled by dropping the corresponding terms from Eqs.~(\ref{prob_spec}) and 
(\ref{prob_spec_special}). Theoretically confirming or disproving that the conditions $K<0$ and $P \le 0$ 
always hold is however out of the scope of current study.

\subsection{The importance of priors}\label{priors_discussion}
\begin{figure}
	\includegraphics[width=0.5\textwidth, trim={0 0.5cm 0.5cm 0.5cm}, clip]{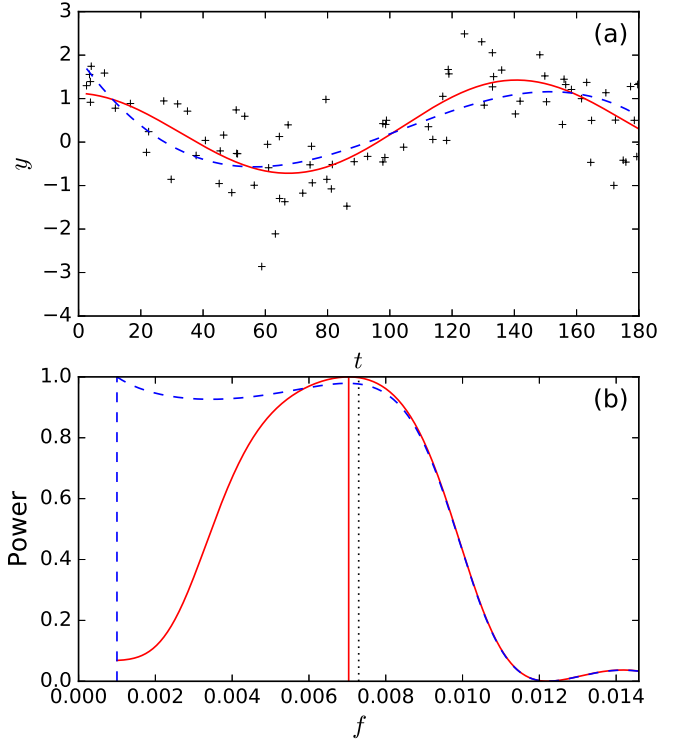}
	\caption{Illustration of importance of priors. (a) Models fitted to the data with Gaussian priors 
	(red continuous line) and uniform priors (blue dashed curve) for the nuisance parameters.
	(b) The spectra of the corresponding models. The black dotted line shows the position of true frequency.
	}
	\label{fig_prior}
\end{figure}

In Eq.~(\ref{eq_prior}) we defined the priors of the nuisance parameters $\bm{\theta}$ to be Gaussian with
reasonable means and variances given in Eq.~(\ref{eq_priors}). In this subsection we discuss the significance
of this choice to the results. The presence of the linear trend component introduces additional degree of 
freedom into the model, which, in the context of low frequencies and short datasets can lead to large and 
physically meaningless parameter values, when no regularization is used. 
This situation is illustrated in Fig.~\ref{fig_prior}, where we show the difference 
between the models with Gaussian and uniform priors  used
for the vector of nuisance parameters $\bm{\theta}$.
The true parameter vector in this example was
$\bm{\theta}=[0.8212, -0.5707, 0.003258, 0]$ and
the estimated parameter vector for the model with Gaussian priors
$\bm{\theta}=[0.7431, -0.6624, 0.002112, 0.1353]$. However,
for the model with uniform priors the estimate 
$\bm{\theta}=[3.016, 98.73, -0.5933, 61.97]$ is strongly deviating from the true vector.
As seen from the Fig.~\ref{fig_prior}(a), from the point of view of the goodness of fit both of these solutions differ 
only marginally. 
From the perspective of parameter estimation (including the period), however, the model with uniform priors leads to substantially biased results.
From Fig.~\ref{fig_prior}(b) it is evident that $p(f|D)$ can become multimodal when 
uniform priors are used and in the worst case scenario the global maximum occurs in the low end of the
frequency range, instead of the neighbourhood of the true frequency.

\subsection{Dealing with multiple harmonics}

The formula for the spectrum is given by Eq.~(\ref{prob_spec}), which represents the posterior probability 
of the frequency $f$, given data $D$ and our harmonic model $M_{\rm H}$ i.e. $p(f|D,M_{\rm H})$. 
Although being a
probability density, it is still convenient to call this frequency-dependent quantity a spectrum. We point out, 
that if the true
model matches the given model, $M_{\rm H}$, then the interpretation of the spectrum is straightforward, namely 
being the probability distribution 
of the frequency. This will give us a direct way for error estimation, e.g. by fitting a Gaussian to the spectral line \citep[Chapter~2]{Bretthorst1988}.
When the true model has more than one harmonic, the interpretation of the spectrum is not anymore direct due to 
mixing of the probabilities from different harmonics.
The correct way to address this issue would be to use more 
complex model with at least as many harmonics that there are expected to be in the underlying process (or even 
better, to infer the number of components from data). However, as a simpler workaround, the ideas of cleaning 
the spectrum introduced in \citet{Roberts1987} can be used to iteratively extract significant frequencies from 
the spectrum calculated using a model with single harmonic. 

\subsection{Significance estimation}

To estimate the significance of the peaks in the spectrum
we perform a model comparison between
the given model and a model without harmonics, i.e. only with linear trend. 
One practical way to do this is to calculate
\begin{equation}\label{delta_bic}
\Delta {\rm BIC}={\rm BIC}_{M_{\rm null}} - {\rm BIC}_{M_{\rm H}},
\end{equation}
where $M_{\rm null}$ is the linear model without harmonic, 
${\rm BIC}=\ln(n)k-2ln(\hat{L})$ is the Bayesian Information Criterion (BIC), $n$ is the number of data 
points, $k$ the number of model parameters and $\hat{L}=p(D|\hat{\theta},M)$ is the likelihood of data for 
model $M$ using the parameter values that maximize the likelihood. This formula is 
an approximation to the logarithmic Bayes factor
\footnote{Bayes factor is the ratio of the marginal likelihoods of the data under two hypothesis (usually null and an alternative).
In frequentists' statistics there is no direct analogy to that, but it is common to calculate the p-value of the test statistic
(e.g. the $\chi^2$ statistic in period analysis). 
The p-value is defined as the probability of observing the test statistic under the null hypothesis to 
be larger than that actually observed \citep[Ch.~5.3.3]{Murphy}.}, more precisely $\Delta {\rm BIC} \approx 2 \ln K$, where $K$ 
is the Bayes factor. Strength 
of evidence for models with $2 \ln K > 10$ are considered very strong, with $6 \le 2 \ln K < 10$ strong, and 
with $2 \le 2 \ln K < 6$ positive. For a harmonic model $M_{\rm H}$, $\theta$ include optimal frequency, and 
coefficients of the harmonic component, slope and offset, for $M_{\rm {null}}$ only the last two coefficients. 

In the derivation of the spectrum we assumed that the noise variance of the data points is known, which is 
usually not the case. In probabilistic models, this parameter should also be optimized,
however in practice often sample variance is taken as the estimate for it. This is also the case with LS 
periodogram. 
However, it has been shown that when normalizing the periodogram with the sample variance the statistical 
distribution slightly differs from the theoretically expected one \citep{Schwarzenberg-Czerny1998},
thus it has an effect on the significance estimation.
More realistic approach, especially in the case when the noise cannot be assumed stationary
is to use e.g. subsample variances in a small sliding window around the data points.

As a final remark in this section, we want to emphasize that the probabilistic approach does not remove the burden 
of dealing with spectral aliases due to data sampling. There is always a chance of false detection, thus the 
interpretation of the spectrum must be done with care \citep[for a good example see][]{Pelt1997}.

\section{Experiments}\label{experiments}
In this section we undertake some experiments to compare the performance of the introduced method with
LS and/or GLS periodograms. 

\subsection{Performance of the method in the absence of a trend}\label{exp_zero_trend}
We start with the situation where no linear trend is present in the actual data,
as this kind of a test will allow us to compare the performance of the BGLST model, with
an additional degree of freedom, to the GLS method.
For that purpose we draw $n$ data points randomly from a harmonic process with zero mean and total variance of unity.
The time span of the data $\Delta T$ is selected to be $30$ units.
We did two experiments, one with uniform sampling and the other with alternating segments of data and gaps (see below).
In the both experiments we varied $n$ in the range from $5$ to $100$ and noise variance $\sigma^2_{\rm N}$ from $0$ to $0.5$. 
The period of the harmonic was uniformly chosen from the range between $0$ and $20$ in the first experiment and from the range between $0$ to $6$ in the second experiment.
As a performance indicator we use the following statistic, 
which measures the average relative error of the period estimates:
\begin{equation}\label{eq_S1}
S_1=\frac{1}{N}\sum_{i=1}^{N}\delta_i,
\end{equation}
where $N$ is the number of experiments with identical setup, $\delta_i=\Delta_i/f^{\rm true}$ and $\Delta_i = |f_i - f^{\rm true}_i|$ is the 
absolute error of the period estimate 
in the $i$-th experiment using the given method (either BGLST or GLS).

\begin{figure}
	\includegraphics[width=0.5\textwidth, trim={3.0cm 0.5cm 1.0cm 1.0cm}, clip]{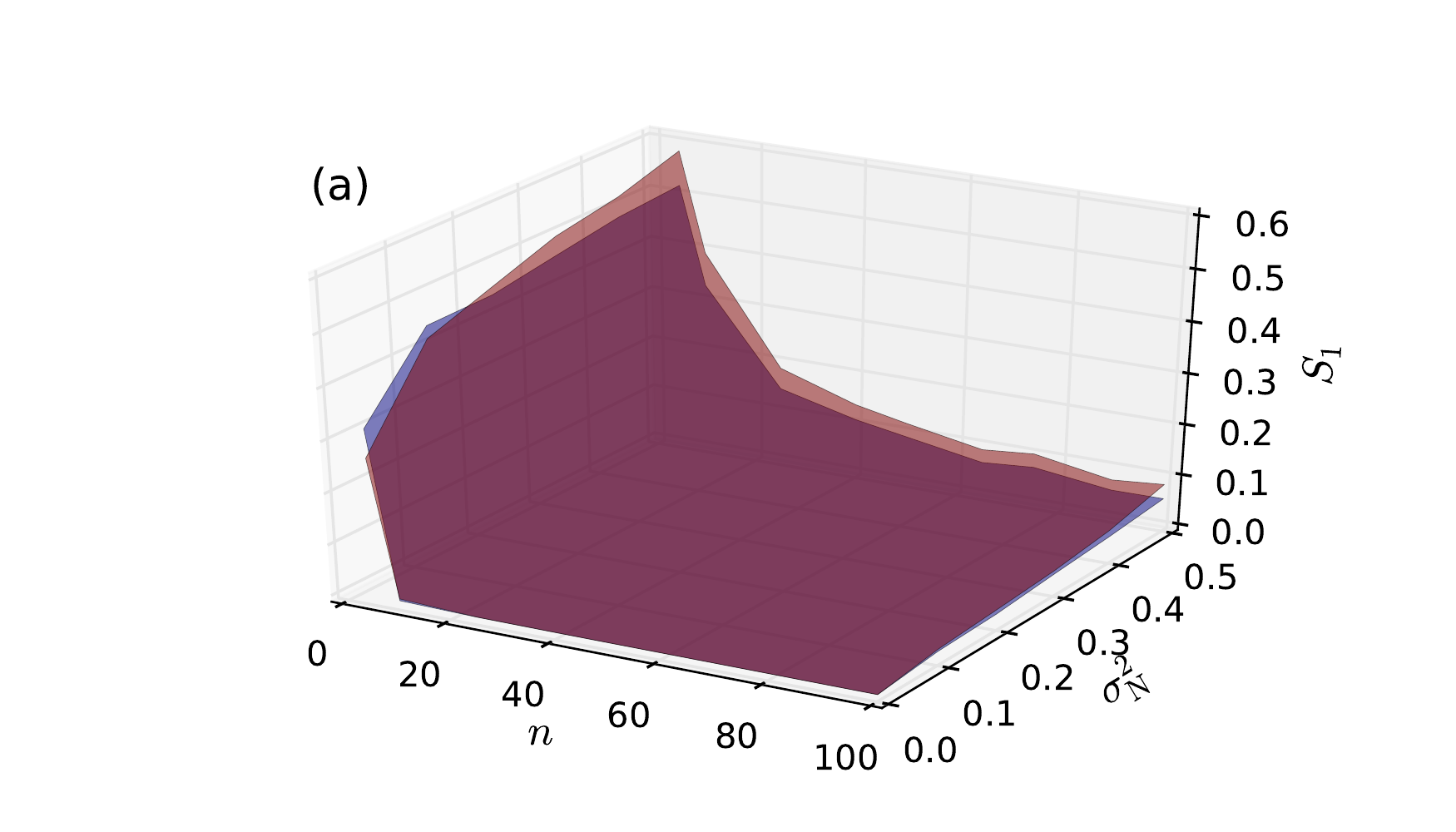}
	\includegraphics[width=0.5\textwidth, trim={3.0cm 0.5cm 1.0cm 1.0cm}, clip]{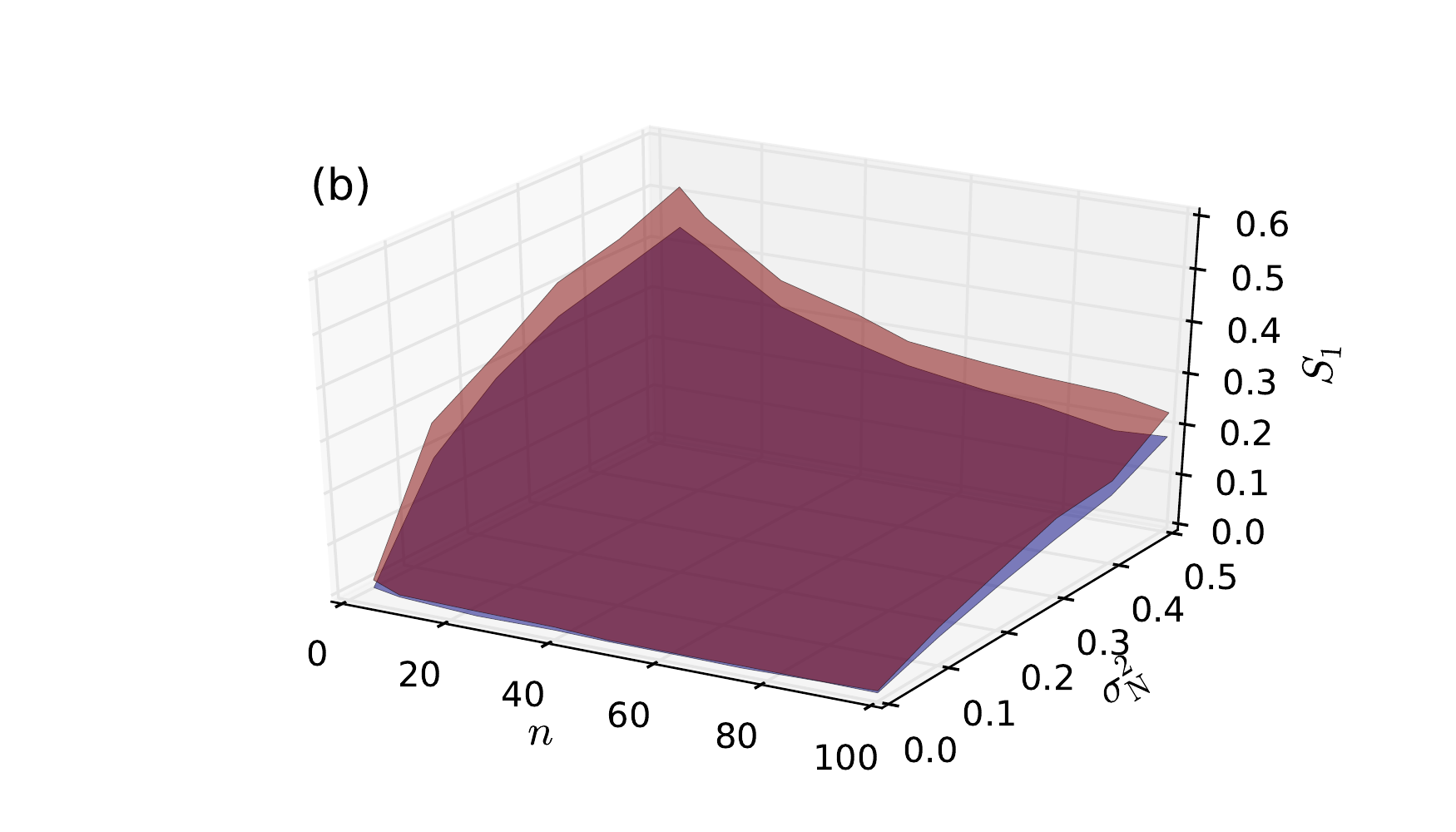}
	\caption{
		The performance measure $S_1$ of BGLST (red) and GLS (blue) methods as function of number of data 
		points $n$ and noise variance $\sigma^2_{\rm N}$. (a) Uniform sampling, (b) sampling with segments and gaps.
		For the definition of $S_1$ see text.
	}
	\label{offset_limit_stats}
\end{figure}

First we noticed that both methods performed practically identically when the sampling was uniform. This result was 
only weakly depending on the number of data points $n$ and the value of the noise variance $\sigma^2_{\rm N}$ (see Fig.~\ref{offset_limit_stats}(a)).
However, when we intentionally created 
such segmented sampling patterns which introduced the presence of the empirical trend, GLS started to 
outperform BGLST more noticeably for low $n$ and/or high $\sigma^2_{\rm N}$. In the latter experiment we created the datasets with 
sampling consisting of five data segments separated by longer gaps. In Fig.~\ref{offset_limit_stats}(b) we show 
the corresponding results. It is clear that in this special setup the performance of 
BGLST gradually gets closer and closer to the performance of GLS when either 
$n$ increases or 
$\sigma^2_{\rm N}$ decreases. However, compared to the uniform case, the difference between the methods is bigger for low $n$ and high $\sigma^2_{\rm N}$.
The existence of the difference in performance
is purely due to the extra parameter in BGLST model, and it is a well known fact that models with higher number of parameters become more prone to overfitting.

\begin{figure}
	\includegraphics[width=0.5\textwidth, trim={0 0.0cm 0.5cm 0.5cm}, clip]{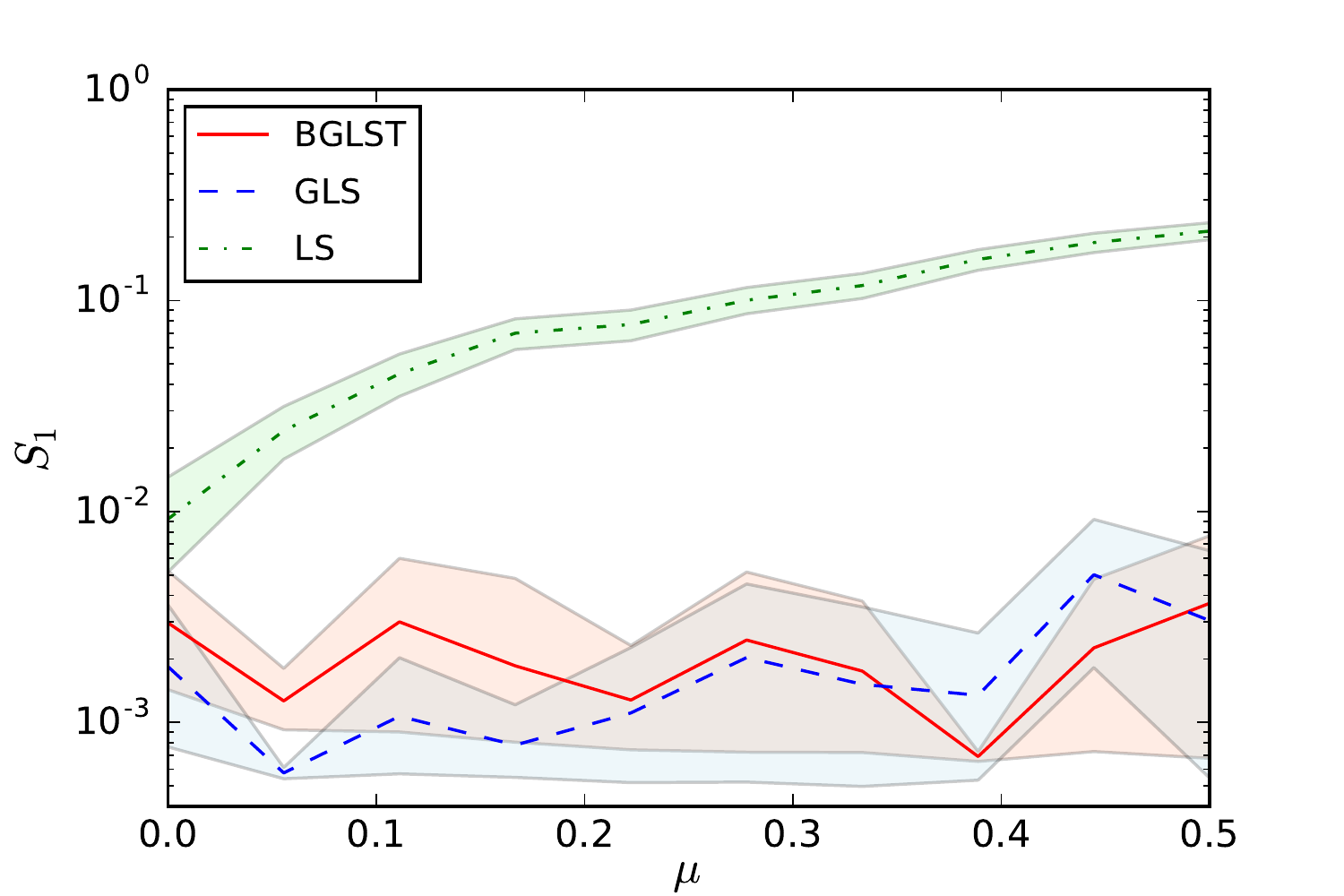}
	\caption{
		The performance of BGLST, GLS and LS as function of sample mean $\mu$. Both the true mean and slope are zero.
		The shaded areas around the curves on this
		and all subsequent plots show the 95\% confidence intervals of the standard error of the statistic.
	}
	\label{offset_stats}
\end{figure}

The effect of an offset in the randomly sampled data to the period estimate has been well described in 
\citet{Mortier2015}. Using GLS or BGLS periodograms instead of LS, one can eliminate the potential bias 
from the period estimates due to the mismatch between the sample mean and true mean.
We conducted another experiment to show the performance comparison 
of the different methods in this situation. We used otherwise identical setup as described in the second experiment 
above, but we fixed $n=25$ to more easily control the sample mean and we also set the noise variance to zero. 
We measured how the performance statistic $S_1$ of the methods changes as function of the sample mean 
$\mu$ (true mean was zero).
The results are shown in Fig.~\ref{offset_stats}.
We see that the relative mean error of the LS method steeply increases with increasing discrepancy between 
the true and sample mean, however both the performance of BGLST and GLS stays constant and relatively 
close to each other. 
Using non-zero noise variance in the experiment increased $S_1$ of BGLST slightly higher than 
GLS due to the effects described in the previous experiments, but the results were still independent of 
$\mu$ as expected.

\subsection{Effect of a linear trend}\label{exp_trend}

\begin{figure}
	\includegraphics[width=0.5\textwidth, trim={0 0.5cm 0.0cm 0.0cm}, clip]{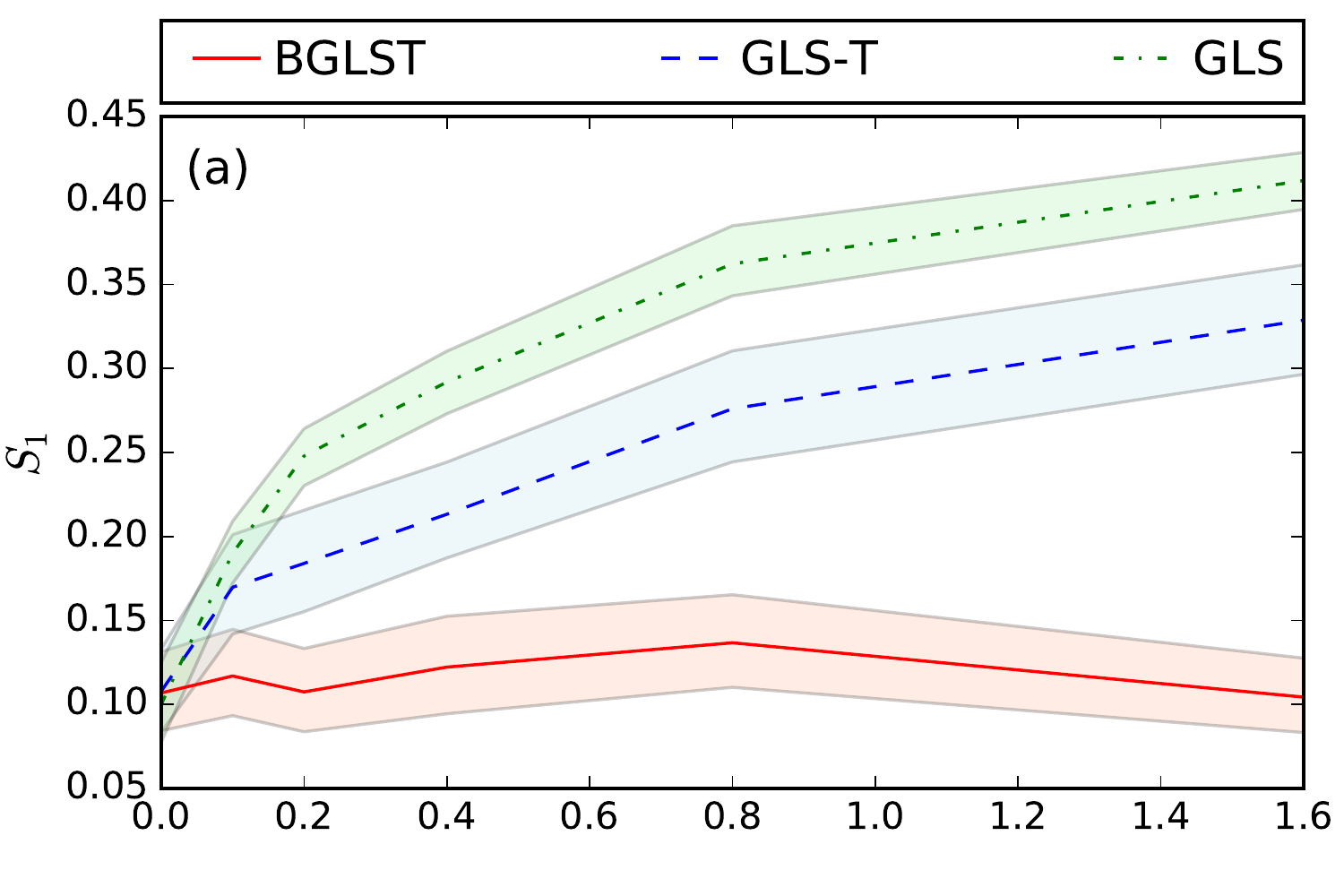}
	\includegraphics[width=0.5\textwidth, trim={0 0.0cm 0.0cm 0.5cm}, clip]{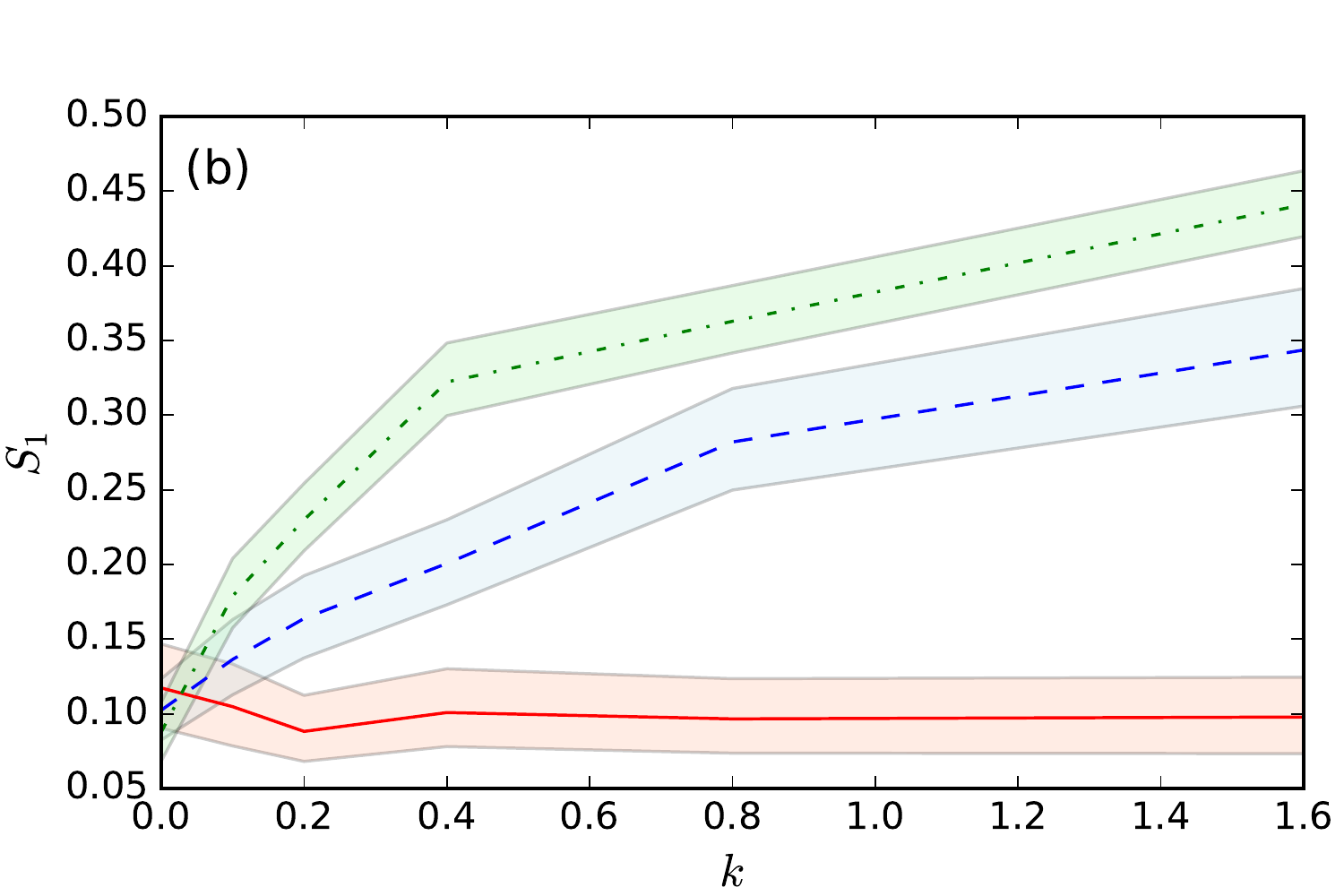}
	\caption{
	Results of the experiments with varying linear trend using 
	uniform sampling (a) and sampling similar to MW datasets (b).
	}
	\label{fig_trend_stats}
\end{figure}

Let us continue with the question of how much the presence of a linear trend in the data affects the period estimate.
To measure the performance of the BGLST method introduced in Sect.~\ref{harmonic_model},
we do the comparison to
plain GLS and GLS with preceding linear detrending (GLS-T). 
Throughout this section we assume a constant noise variance. 
We draw the data with time span $\Delta T \approx 30$ time units randomly from the 
harmonic process with a linear trend
\begin{equation}
y(t)=A\cos(2\pi f t) + B\cos(2\pi f t) + \alpha t + \beta + \epsilon(t),
\end{equation}
where $A$, $B$, $\alpha$ and $\beta$ are zero mean independent Gaussian random variables with
variances $\sigma^2_{\rm A} = \sigma^2_{\rm B} = \sigma^2_{\rm S}$, $\sigma^2_{\alpha}$ and $\sigma^2_{\beta}$ respectively, and
$\epsilon(t)$ is a zero mean Gaussian white noise process with variance $\sigma^2_{\rm N}$.
We varied the
trend variance $\sigma^2_{\rm \alpha}=k (\sigma^2_{\rm S} + \sigma^2_{\rm N}) /\Delta T$, such that 
$k \in [0, 0.1, 0.2, 0.4, 0.8, 1.6]$. 
For each $k$ we generated $N=2000$ time series, where each time
the signal to noise ratio (SNR=$\sigma^2_{\rm S}/\sigma^2_{\rm N}$) was drawn from
$[0.2, 0.8]$ and period $P=1/f$ from $[2, 2/3\Delta T]$. In all experiments $\sigma^2_{\beta}$ was set equal to $\sigma^2_{\rm S} + \sigma^2_{\rm N}$.
We repeated the 
experiments with two forms of sampling: uniform and the one based
on the samplings of MW datasets.
The prior means and variances for our model were chosen
according to Eq.~(\ref{eq_priors}).

In these experiments we compare three different methods -- the BGLST, GLS-T and GLS.
We measure the performance of each method
using the statistic $S_1$ defined in Eq.~(\ref{eq_S1}).

The results for the experiments with the uniform sampling are shown in 
Fig.~\ref{fig_trend_stats}(a) and with the sampling taken from the MW dataset in 
Fig.~\ref{fig_trend_stats}(b). 
On both of the figures we plot the performance measure $S_1$ as function of $k$.
We see that
when there is no trend present in the dataset ($k=0$), all three methods have approximately 
the same average relative errors
and while for the BGLST method it stays the same or slightly decreases with increasing $k$, for
the other methods the errors start to increase rapidly. We also see that when the true trend 
increases then GLS-T starts to outperform GLS, however, the performance of both of these 
methods stays far behind from the BGLST method.

\subsubsection{Special cases}

\begin{figure}
	\includegraphics[width=0.5\textwidth, trim={0 0.5cm 0 1cm}, clip]{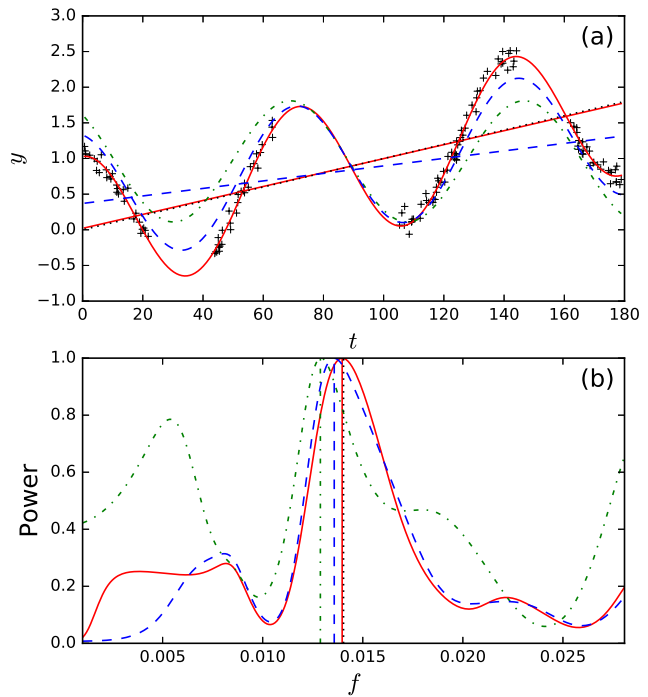}
	\caption{Comparison of the results using different models. The true model of the data 
	contains one harmonic, trend and additive white Gaussian noise. (a) Data (black crosses), 
	BGLST model (red continuous curve), GLS-T model (blue dashed curve), 
	GLS model (green dash-dotted curve), true 
	trend (black dotted line), trend from BGLST model (red continuous line), 
	empirical trend (blue dashed 
	line). (b) Spectra of the 
	corresponding models with vertical lines marking the locations of maxima.
	The black dotted line shows the position of the true frequency.
	}
	\label{fig_model_comp}
\end{figure}

Next we illustrate the benefit of using BGLST model with two examples, where the
differences to the other models are well emphasized. 
For that purpose we first generated such a dataset where the empirical slope 
significantly differs from the true slope. The test contained only one harmonic with a
frequency of 0.014038 and a slope 0.01.
We again fit three models -- BGLST, GLS and GLS-T. 
The comparison of the results are shown in Fig.~\ref{fig_model_comp}. 
As is evident from Fig.~\ref{fig_model_comp}(a), the empirical trend 0.0053 recovered
by the GSL-T method (blue dashed line) differs significantly from the true trend
(black dotted line). Both the GSL and GSL-T methods perform very poorly in fitting
the data, while only the BGLST model produces a fit that represents the data points
adequately. 
Moreover, BGLST recovers very close to the true trend value 0.009781.
From Fig.~\ref{fig_model_comp}(b) it is evident that the BGLST model
retrieves a frequency closest to the true one, while the other two methods return
too low values for it. The performance of the GLS model is the worst, as expected.
The frequency estimates for BGLST, GLS-T and GLS are correspondingly: 
0.013969, 0.013573 and 0.01288.
This simple example shows that
when there is a real trend in the data, due to the 
sampling patterns, detrending can be erroneous, but it still leads to better results than not 
doing the detrending at all.

\begin{figure}
	\includegraphics[width=0.5\textwidth, trim={0 0.5cm 0 1cm}, clip]{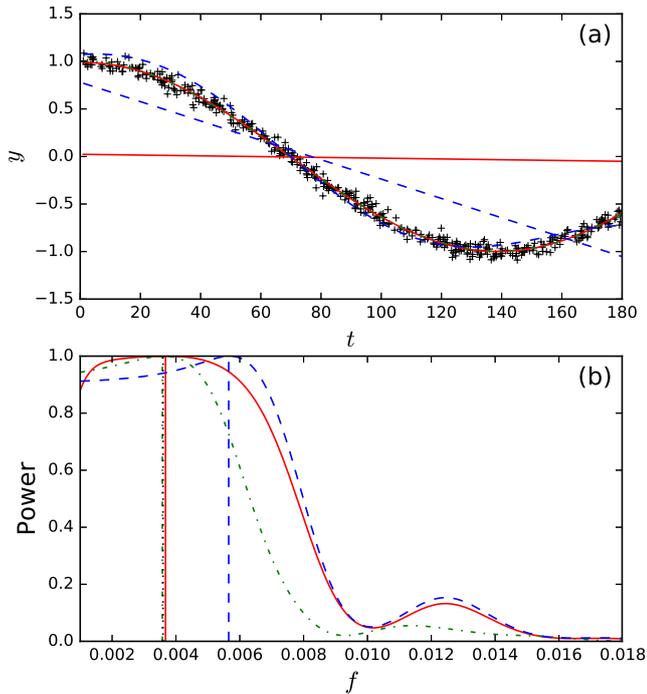}
	\caption{Comparison of the results using different models. The true model contains one long 
	harmonic with additive white Gaussian noise. (a) Data (black crosses), BGLST model
	(red continuous curve) and its trend component (red line), GLS-T model with trend added back (blue 
	dashed curve), empirical trend (blue dashed line), GLS model 
	(green dash-dotted curve). (b) Spectra of the corresponding models with vertical lines 
	marking the locations of maxima.
	The black dotted line shows the position of the true frequency.
	}
	\label{fig_model_comp_2}
\end{figure}

Finally we would like to show a counterexample where detrending is 
not the preferred option. This happens when there is a harmonic signal in the data with a very long 
period,
in this test case 0.003596.
The exact situation is shown in Fig.~\ref{fig_model_comp_2}(a),
wherefrom it is evident that the GLS-T method can determine the harmonic
variation itself as a trend component, and lead to a completely erroneous
fit.
In Fig.~\ref{fig_model_comp_2}(b) we show the corresponding spectra. As expected, the GLS model,
coinciding with the true model, gives the 
best estimate, while the BGLST model is not far off. 
The detrending, however, leads to significantly worse estimate. 
The estimates in the same order are the following: 0.003574, 0.003673 and 0.005653.
The value of the trend learned by BGLST method was -0.00041, which is very close to zero.

The last example clearly shows that even in the simplest case of pure harmonic, if the 
dataset does not contain exact number of periods the spurious trend component arises. 
If pre-detrended, then bias is introduced into the period estimation, however,
in the case of BGLST method the sinusoid can be fitted into the fragment of the harmonic
and zero trend can be recovered.

\subsection{Effect of a nonconstant noise variance}\label{exp_noise}

We continue with investigating the effect of a nonconstant noise variance to the period estimate. 
In these experiments the data is generated from a purely harmonic model with
no linear trend (both $\alpha$ and $\beta$ are zero).
We compare the results 
from the BGLST
to the LS method.
As there is no slope and offset in the model, we set the prior means for these two parameters
to zero and variances to very low values. This essentially means that we are approaching
the GLS method
with a zero mean. 
In all the experiments the time range of the observations is $t_i \in [0, T]$, 
where $T=30$ units, the period is drawn from $P \sim \rm{Uniform}(5, 15)$ and 
the signal variance is $\sigma^2_{\rm S}=1$. For each experiment we draw two 
values for the noise variance  $\sigma^2_1 \sim \rm{Uniform}(2\sigma^2_{\rm S}, 10\sigma^2_{\rm S})$ and 
$\sigma^2_2 = \sigma^2_1/k$, where $k \in [1, 2, 4, 8, 16, 32]$ which are used
in different setups as indicated in the second column of Table~\ref{tab_exp_noise}. For each $k$ we repeat the experiments $N=2000$ times.

Let $\Delta_i = |f_i - f^{\rm true}_i|$ denote the absolute
error of
the BGLST
period estimate in the $i$-th experiment and $\Delta^{\rm LS}_i = |f^{\rm LS}_i - f^{\rm true}_i|$ the 
same for the LS method.
Now denoting by $\delta_i=\Delta_i/f^{\rm true}$ and $\delta^{\rm LS}_i=\Delta^{\rm LS}_i/f^{\rm true}$ the relative errors of the corresponding period estimates,
we measure the following performance statistic, which represents the relative difference in the average 
relative errors between the methods i.e.
\begin{equation}\label{eq_S2}
S_2=1-\frac{\sum_{i=1}^{N}\delta_i}{\sum_{i=1}^{N}\delta^{\rm LS}_i}.
\end{equation}

The list of experimental setups with the descriptions of the models are shown in 
Table.~\ref{tab_exp_noise}.
In the first experiment the noise variance is linearly increasing from $\sigma^2_1$ to 
$\sigma^2_2$. In practice this could correspond to decaying measurement accuracy over time. In the second experiment 
the noise variance abruptly jumps from $\sigma^2_2$ to $\sigma^2_1$ in the middle of the
time series. This kind of situation could be interpreted as a change of one instrument
to another, more accurate, one.
In both of these experiments the sampling is uniform. In the third experiment we use sampling patterns 
of randomly chosen stars from MW dataset and the 
true noise variance in the generated data is set based on the empirical intra-seasonal variances in the real data.
In the first two experiments the number of data points was $n=200$ and in the third between $200$ and $400$
depending on the randomly chosen dataset, which we downsampled.

\begin{table}\caption{Description of experimental setups with nonconstant noise variance.
The first column indicates the number of the setup, the second column shows how
the variance $\sigma^2_i$ for $i$-th data point was selected and third column the 
criteria of drawing the time moment $t_i$ for $i$-th data point.
		}
	\begin{tabular}{c|c|c}
		No. & Form of noise & Type of sampling \\ \hline
		1 & $\sigma^2_i = \sigma^2_1 + 
		\frac{t_i}{T}(\sigma^2_2-\sigma^2_1)$ 
		& $t_i \sim \rm{Uniform}(0, T)$ \\
		2 & $\sigma^2_i = \begin{cases}
		\sigma^2_2, \text{if } t_i < T/2 \\
		\sigma^2_1, \text{if } t_i \ge T/2
		\end{cases}$                     & $t_i \sim \rm{Uniform}(0, T)$ \\
		3 & \begin{tabular}{@{}l@{}}$\sigma^2_i = \sigma^2_2 + $\\$\frac{\sigma^{2\rm ({emp})}_i-\sigma^{2\rm ({emp})}_{\rm {min}}}{\sigma^{2\rm ({emp})}_{\rm {max}}-\sigma^{2\rm ({emp})}_{\rm {min}}}(\sigma^2_1-\sigma^2_2)$ \end{tabular}& Based on MW dataset \\
	\end{tabular}
	\label{tab_exp_noise}
\tablefoot{
In the \nth{3} row the $\sigma^{2\rm ({emp})}_i$ denotes the empirical 
variance of the $i$-th datapoint and $\sigma^{2\rm ({emp})}_{\rm {min}}$ and $\sigma^{2\rm ({emp})}_{\rm {max}}$
the minimum 
and maximum intra-seasonal variances. In other words we have renormalized the intra-seasonal variances to the interval from
$\sigma^2_1$ and $\sigma^2_2$.
	}	
\end{table}

\begin{figure}
	\includegraphics[width=0.5\textwidth, trim={0 0.5cm 0.0cm 0.0cm}, clip]{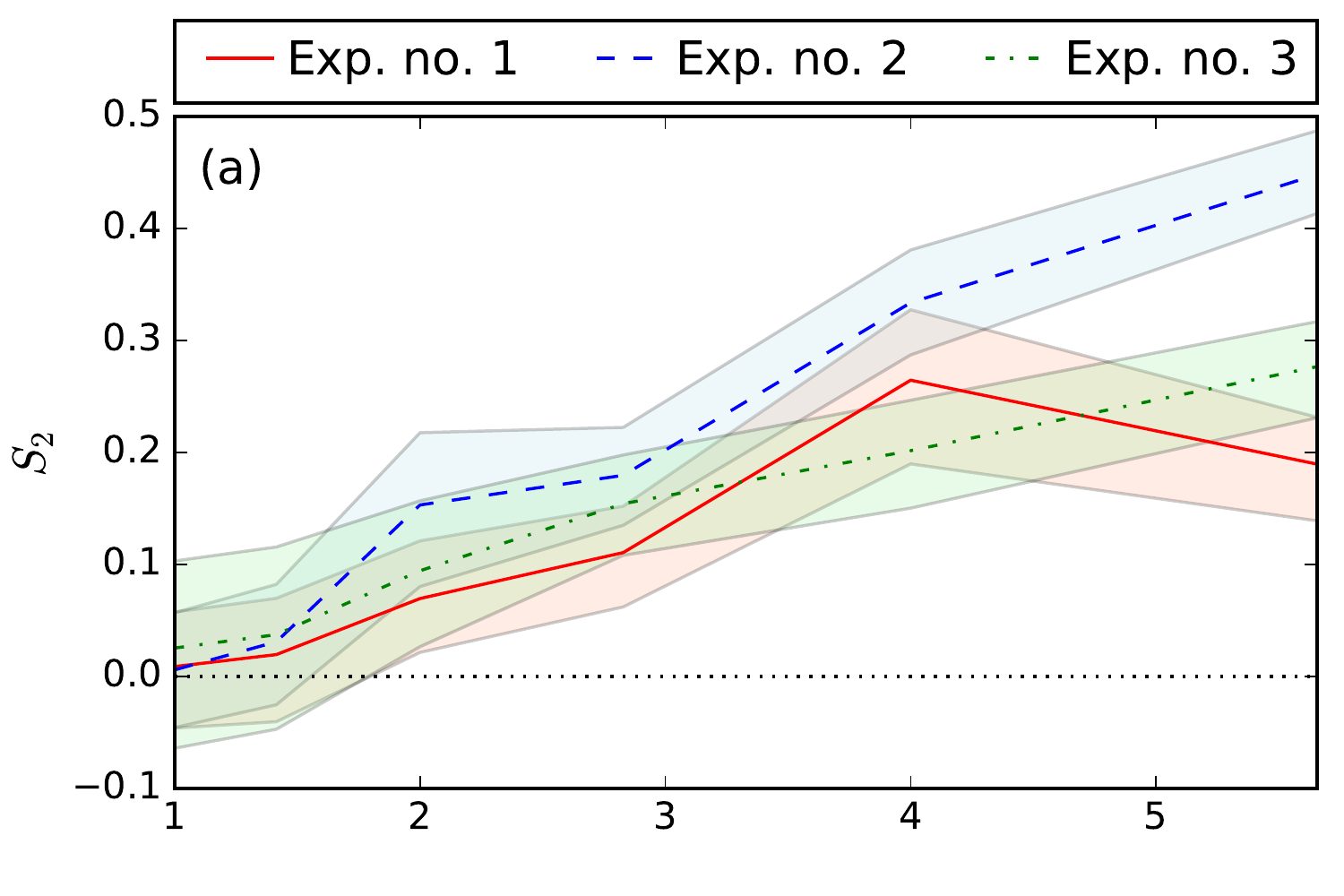}
	\includegraphics[width=0.5\textwidth, trim={0 0.0cm 0.0cm 0.5cm}, clip]{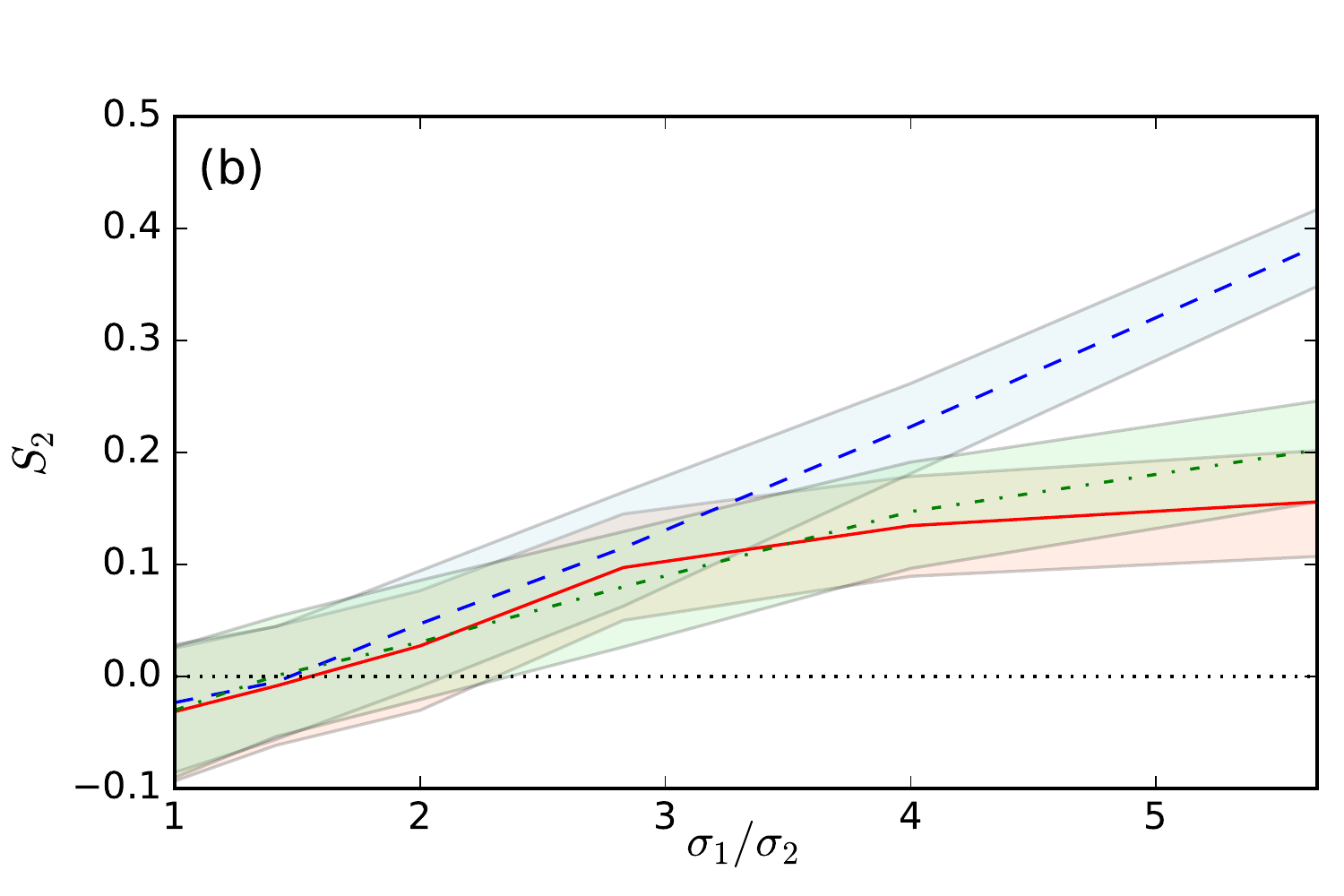}
	\caption{Results of the experiments with known nonconstant noise variance. 
		For the definition of $S_2$ see text.
		(a) True noise variance is known. (b) Noise variance is empirically estimated from the data.
		The black dotted horizontal lines show the break even point between LS and BGLST.
	}
	\label{fig_noise_stats}
\end{figure}

In Fig.~\ref{fig_noise_stats}(a) the results of different experimental setups are shown, where 
the performance statistic $S_2$ is plotted as function of $\sqrt{k}=\sigma_1/\sigma_2$.
We see that when the true noise is constant in the data ($k=1$), the BGLST method performs
identically to the LS method while for greater values of $k$ the difference grows
bigger between the methods. For the \nth{2} experiment, where the noise level abruptly
changes, the advantage of using a model with nonconstant noise seems to be the best,
while for the other experiments the advantage is slightly smaller, but still substantial for larger $k$-s.

\begin{figure*}
	\includegraphics[width=1.0\textwidth, trim={0 0.5cm 0 1cm}, clip]{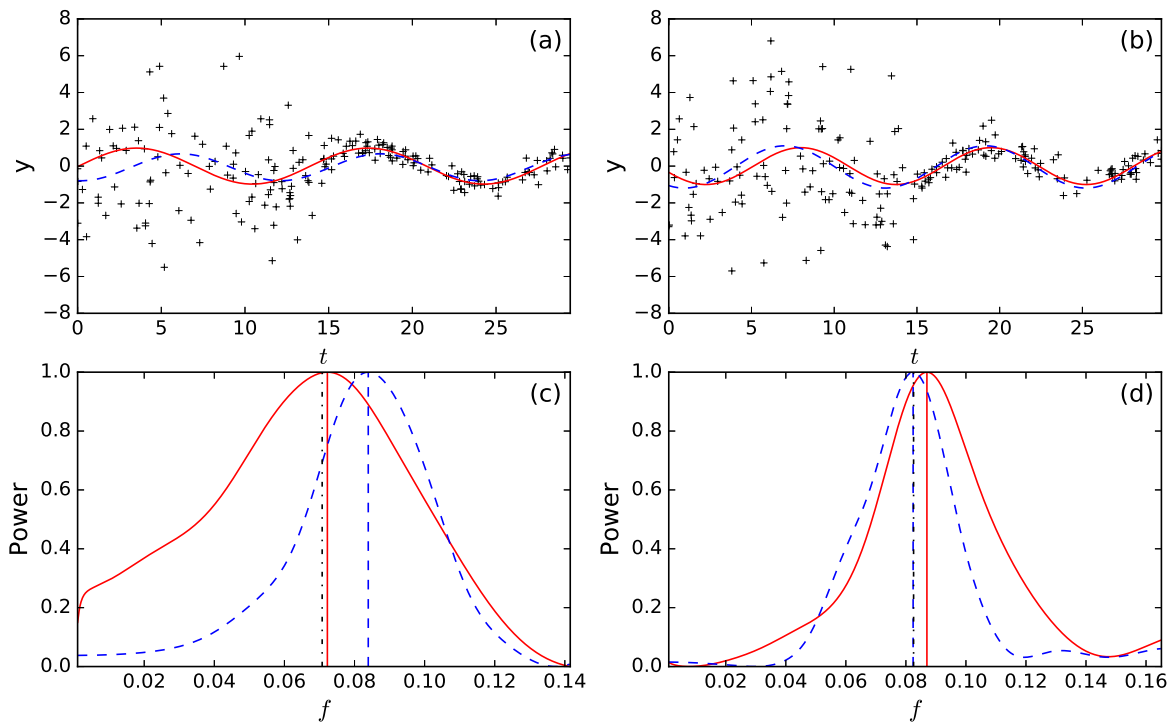}
	\caption{Comparison of the methods for experiment with setup 2.
		(a) and (b) Data points with black crosses,
		red continuous curve -- BGLST model, blue dashed line -- LS model fit. (c) and (d) Spectra of the
		models with same colours and line styles. The vertical lines correspond to the optimal frequencies found
		and the black dotted line shows the true frequency. The plots on two columns correspond to the biggest difference in the period
		estimates from both methods w.r.t. each other. On the left BGLST method outperforms LS method, on the right vice versa.		
	}
	\label{fig_noise_exp_2}
\end{figure*}

In Fig. \ref{fig_noise_exp_2} we have shown the models 
with maximally differing period estimates from the \nth{2} experiment for $\sqrt{k}=5.66$. On the left column of the 
figures there is shown the situation where BGLST method outperforms the LS 
method the most and on the right column vice versa. For colour and symbol
coding, please see the caption of Fig.~\ref{fig_noise_exp_2}. 
This plot is a clear illustration of the fact that even though on average the method 
with nonconstant noise variance is better than LS method, for each particular dataset 
this might not be the case. Nevertheless it is also apparent from the figure that when
the LS method is winning over BGLST method, the gap tends to be slightly smaller than
in the opposite situation.

In the previous experiments the true noise variance was known, but this rarely happens in practice. However, when 
the data sampling is sufficiently dense, we can estimate the noise variance empirically 
e.g. by binning the data using a window with suitable width.
We now repeat the experiments using such an approach.
In experiment setups 1 and 2 we used windows with 
the length 1 unit and in the setup 3 the intra-seasonal variances. In the first two cases we increased the number 
of data points to 1000 and in the latter case we used only those real datasets which contained more than 500 points. 
The performance statistics for these
experiments are shown in Fig.~\ref{fig_noise_stats}(b). 
We see that when the true noise variance is constant ($k=1$)
then using empirically estimated noise in the model leads to slightly worse results than with constant noise model,
however, roughly starting from the value of $k=1.5$ in all the experiments the former approach starts to outperform
the latter.
The location of the break-even point obviously depends on how precisely the true variance can be estimated 
from the data. This, however depends on how dense is the data sampling and how short is the expected period in the data, 
because we want to avoid counting signal variance as a part of the noise variance estimate.

\subsection{Real datasets}

\begin{figure*}
	\includegraphics[width=1.0\textwidth, trim={0 0.5cm 0 0.5cm}, clip]{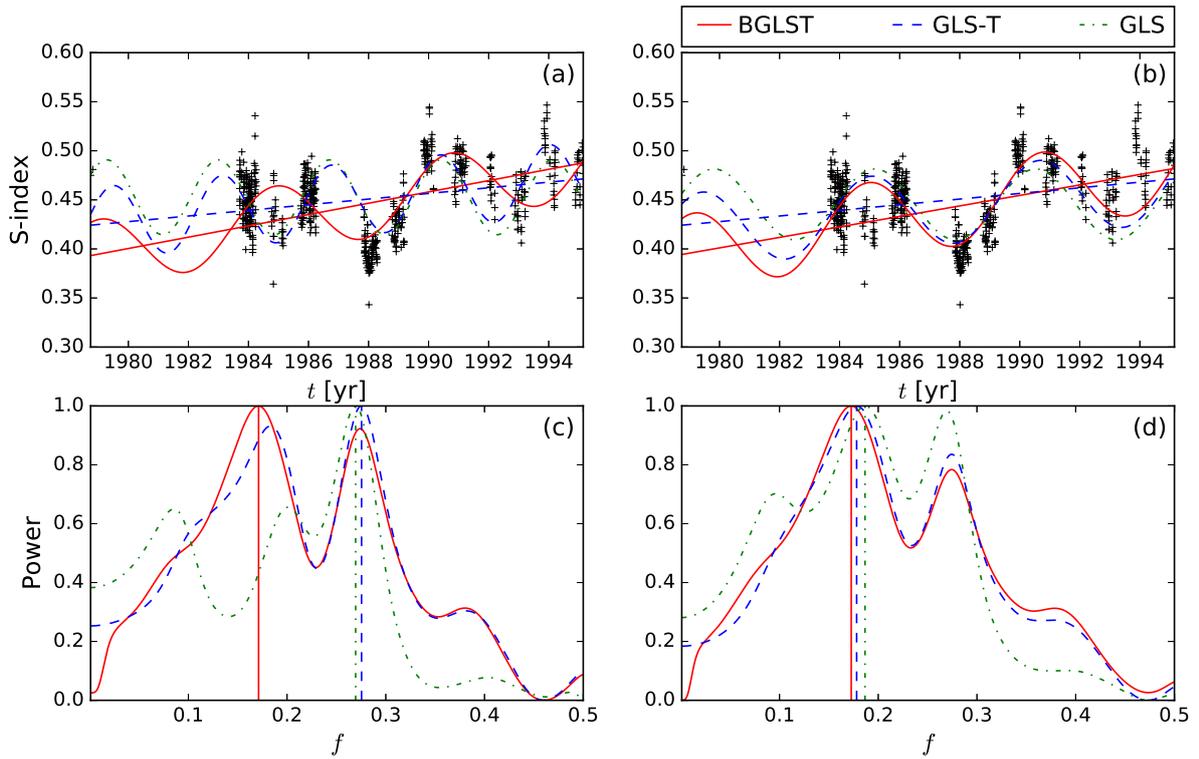}
	\caption{Comparison of the results for star HD37394 using BGLST, GLS-T and GLS models with constant noise 
		variance in the plots on the left column and with intra-seasonal noise variance on the right column.
		(a) and (b) Data (black crosses), BGLST model 
		(red curve), GLST-T model with trend added back (blue dashed curve), GLS model (green dash-dotted curve), the trend component 
		of BGLST model (red line) and the empirical trend (blue dashed line). (c) and (d) Spectra of the models with same colours. The dashed  lines mark the locations of the corresponding maxima.}
	\label{fig_model_comp_hd37394}
\end{figure*}

\begin{figure*}
	\includegraphics[width=1.0\textwidth, trim={0 0.5cm 0 0.5cm}, clip]{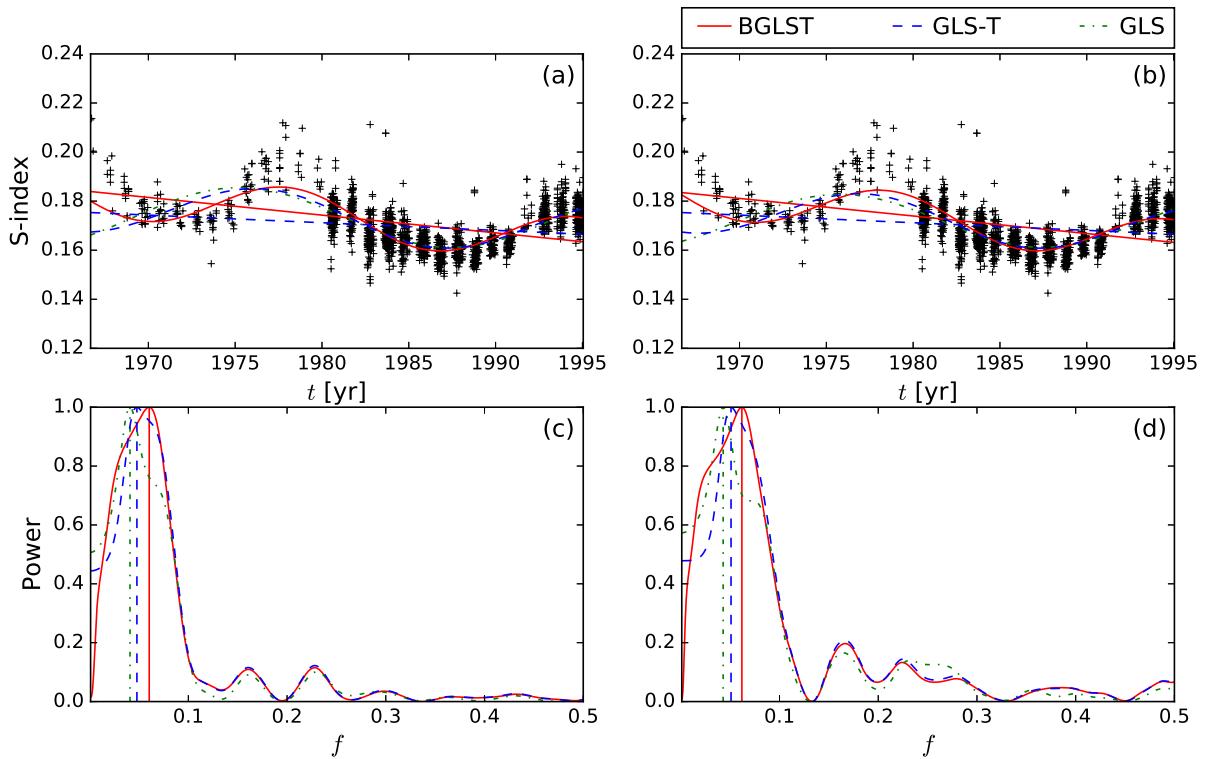}
	\caption{Comparison of the results for star HD3651 using BGLST, GLS-T and GLS models. The meaning of the panels and 
		colour coding is identical to the Fig.~\ref{fig_model_comp_hd37394}.}
	\label{fig_model_comp_hd3651}
\end{figure*}

As the last examples we consider time series of two stars from the MW dataset.
In Figs.~\ref{fig_model_comp_hd37394} and ~\ref{fig_model_comp_hd3651} we show
the differences between period estimates for the stars HD37394 and HD3651 correspondingly.
For both stars we see that the linear trends 
fitted directly to the data (blue dashed lines) significantly differ from the trend component 
in the harmonic model (red lines)
with both types of noise models (compare the left and right columns of the plots), and consequently,
also the period estimates in between the methods always tend to vary somewhat.

Especially in the case of HD37394, the retrieved period estimates differ significantly.
Assuming a constant noise variance for this star results in the BGLST and GLS/GLS-T to deviate
strongly, the BGLST model producing a significantly larger period estimate.
With intra-seasonal noise variance, all
three estimates agree more to each other, however, BGLST still gives the longest period estimate.
In the previous study by \citet{Baliunas1995} the cycle period for this star has been reported to be 3.6~yrs which closely
matches the period estimate of GLS-T estimate with constant noise variance\footnote{To be more precise, in their study they 
	used LS with centering plus detrending, but this does not lead to measurable
	difference in the period estimate compared to GLST-T for the given datasets. We also note that the datasets used in \citet{Baliunas1995} do not span until 1995, but until 1992 and for the star HD37394 they 
	have dropped the data prior 1980. However, here we do not want to stress on the exact comparison between the results, but
	rather show that when applied to the same dataset, the method used in their study can lead to different results from the 
	method introduced in our study.
	}.
For the GLS method the period estimates using constant and intra-seasonal noise variances were correspondingly 
$3.71 \pm 0.02$ and $5.36 \pm 0.05$~yrs,
for the GLS-T the values were $3.63 \pm 0.02$ and $5.61 \pm 0.05$~yrs
and for the BGLST model
$5.84 \pm 0.07$ and $5.79 \pm 0.05$~yrs.
All the given error estimates here correspond to $1\sigma$ ranges of the frequency estimates.

In the case of HD3651 the period estimates of the models are not that sensitive to the chosen noise model, but
the dependence on how the trend component is handled, is significant. Interestingly, none of these period estimates
match the estimate from \citet{Baliunas1995} (13.8~yrs). 
For the GLS method the period estimates using constant and intra-seasonal noise variances were correspondingly 
$24.44 \pm 0.29$ and $23.30 \pm 0.32$~yrs,
for the GLS-T the values were $20.87 \pm 0.29$ and $19.65 \pm 0.34$~yrs
and for the BGLST model
$16.56 \pm 0.23$ and $16.16 \pm 0.14$~yrs.
As can be seen from the model fits in Fig.~\ref{fig_model_comp_hd3651}(a) and (b), they significantly
deviate from the data,
so we must conclude that the true model is not likely to be harmonic. 
Conceptually the generalization of the model proposed in the current paper to the nonharmonic case is straightforward, but 
becomes analytically and numerically less tractable. 
Fitting periodic and quasi-periodic models to the MW data, which are based on Gaussian Processes,
are discussed in \citet{Olspert2017}.

\section{Conclusions}\label{conc}
In this paper we introduced a Bayesian regression model which involves, besides harmonic 
component also a linear trend component with slope and offset.
The main focus of this paper was on addressing the effect of linear trends in data to the period estimate. We showed that when
there is no prior information on whether and to what extent the true model of the data contains linear trend, it is more preferable to include the
trend component directly to the regression model rather than either detrend the data or leave the data undetrended before 
fitting the periodic model. 

Note that one can introduce the linear trend part also directly to GLS model and use the least squares method to solve
for the regression coefficients. However, based on the discussion in Sect.~\ref{priors_discussion} one should consider adding 
L2 regularisation to the cost function, i.e. use the ridge regression. This corresponds to the Gaussian priors in the Bayesian approach.
While using the least squares approach can be computationally less demanding, the main benefits of the Bayesian approach are
the direct interpretability of the spectrum as being proportional to a probability distribution and more straightforward error and 
significance estimations.

In the current work we used Gaussian independent priors for the nuisance parameters $\bm{\theta}$ and uniform prior for the frequency 
parameter $f$. As mentioned, the usage of priors in our context is solely for the purpose of regularisation. However, if one
has actual prior information about the parameters (the shapes of the distributions, possible dependencies between the parameters, the expected locations of higher probability mass, etc), this could be incorporated into the model by using suitable joint distribution.
In principle one could consider also different forms of priors than independent Gaussian for $\bm{\theta}$, but then the problem becomes 
analytically intractable.
In these situations one should use algorithms like Markov Chain Monte Carlo (MCMC) to sample the points from the posterior distribution. 
The selection of uniform prior for $f$ was also made solely for the reason to not lose the tractability.
If one has prior knowledge about $f$, one could significantly narrow down the
grid search interval. This idea was actually used in the current study as we were focusing only on the signals with 
low frequency periods. However, in the most general intractable case one should still rely on the MCMC methods.

We also showed that if the true noise variance of the data is far from being constant, then
by neglecting this knowledge,
period estimates start to deteriorate. Unfortunately in practice the true noise variance is almost never known and in many
cases impossible to estimate empirically (e.g. sparse sampling rate compared to the true frequency). In this study we, however,
focused only on the long period search task,
which made estimating the noise variance on narrow local subsamples possible.
Based on the experiments we saw that if the true noise is not constant and the extremes of the SNR differ at 
least two times, then using a model with empirically estimated noise variance is well justified.

\begin{acknowledgements}
  This work has been supported by the Academy of Finland Centre of
  Excellence ReSoLVE (NO, MJK, JP), Finnish Cultural Foundation grant no. 00170789 (NO)
  and Estonian Research Council (Grant IUT40-1; JP).
  MJK and JL acknowledge the 'SOLSTAR' Independent Max Planck Research Group funding.
\\
\\
The HK\_Project\_v1995\_NSO data derive from the Mount Wilson Observatory HK Project, which was supported by both public and private funds through the Carnegie Observatories, the Mount Wilson Institute, and the Harvard-Smithsonian Center for Astrophysics starting in 1966 and continuing for over 36 years.  These data are the result of the dedicated work of O. Wilson, A. Vaughan, G. Preston, D. Duncan, S. Baliunas, and many others.  
\end{acknowledgements}

\bibliographystyle{aa}
\bibliography{paper}

\end{document}